\documentclass[a4paper, 12pt]{article}
\usepackage[utf8]{inputenc}
\usepackage[english]{babel}
\usepackage{color,amssymb,amsmath, amsthm,graphicx, wrapfig,listings,caption,multicol,float,enumitem,hyperref,algorithm, algorithmic, booktabs}
\usepackage[dvipsnames]{xcolor}
\usepackage[top=1cm, bottom=2cm, left=1cm, right=1cm]{geometry}
\usepackage[makeroom]{cancel}
\begin{document}
\title{A more efficient way of finding Hamiltonian cycle}
\author{Paweł Kaftan}
\maketitle
\section{Introduction}
Algorithm tests if a Hamiltonian cycle exists in
directed and undirected graphs, if it exists - algorithm can often show found Hamiltonian cycle. If you want to test an undirected
graph, such a graph should be converted to the form of directed graph. Algorithm's goal is to solve NP-complete problems of Directed Hamiltonian cycle and Undirected Hamiltonian cycle in polynomial time. Previously known algorithm solving Hamiltonian cycle
problem - brute-force search can't handle relatively small graphs. Algorithm
presented here is referred to simply as "algorithm" in this paper.\\\\
\textbf{Why algorithm is more efficient than brute-force search?}\\\\
\noindent
In order to find Hamiltonian cycle, algorithm should find edges that creates a Hamiltonian cycle. Higher number of edges creates more possibilities to check to solve the problem. Both brute-force search and algorithm use recursive depth-first search. The reason why brute-force search often fails to solve this problem in reasonable amount of time is too large number of possibilities to check in order to solve the problem.\\
Algorithm rests on analysis of original graph and opposite graph to it. Algorithm prefers ''to think over'' which paths should be checked than check many wrong paths.  Algorithm is more efficient than brute-force search because it can:
\begin{itemize}
\item \textbf{remove unnecessary edges from graph} (\ref{UN} \ref{add} \ref{ALN} \ref{LNE} \ref{recursion} \ref{SingleEdge} \ref{Alley}  \ref{Never1HC_RE} \ref{FECR})
\item \textbf{test when Hamiltonian cycle can't exist in graph}(\ref{NotEnoughUN} \ref{LN} \ref{cykl} \ref{GraphDisconnected} \ref{Never1HC_TLD} \ref{Never1HC_SE})
\item \textbf{choose most optimal path(\ref{CE} \ref{5})} 
\end{itemize}
\section{Definitions}
\begin{enumerate}
\item \textbf{Graph} - set of vertices $v_1, v_2$, \dots,  $v_N$\\ and edges  $(a_1,a_2)$, $(a_3,a_4)$, ..., $(a_{M-1},a_M)$ where \{$a_1,a_2$, \dots,  $a_M\}  \in$ $\{v_1, v_2$, \dots,  $v_N\}$ 
\item \textbf{Path} - set of edges $(v_1,v_2)$, $(v_2,v_3)$, \dots , $(v_{K-1},v_K)$ where $v_{1} \ne v_{2} \ne \dots \ne v_{K}$  and graph contains edges $(v_1,v_2)$, $(v_2,v_3)$, \dots , $(v_{K-1},v_K)$
\item \textbf{Cycle} - Path from graph which contains an edge $(v_K, v_1)$ where $v_K$ is last vertex in cycle and $v_1$ is first vertex in cycle.
\item \textbf{Hamiltonian cycle} - cycle of length equal to the number of vertices in graph
\item \textbf{Hamiltonian graph} - graph which contains Hamiltonian cycle
\item \textbf{Vertex degree} - Vertex $v_1$ has degree equal to $W$ if graph contains following $W$ edges $(v_1,b_1)$, $(v_1,b_2)$, \dots , $(v_1,b_W)$
\item \textbf{Opposite graph} - Graph $G$ is set of vertices $v_1, v_2$, \dots,  $v_N$\\ and edges  $(a_1,a_2)$, $(a_3,a_4)$, ..., $(a_{M-1},a_M)$.\\ Graph opposite to $G$ is set of vertices $v_1, v_2$, \dots,  $v_N$\\ and edges  $(a_2,a_1)$, $(a_4,a_3)$, ..., $(a_M,a_{M-1})$
\\\\Example:
\begin{figure}[ht]
\begin{minipage}[b]{0.45\linewidth}
\centering
\includegraphics[width=\textwidth]{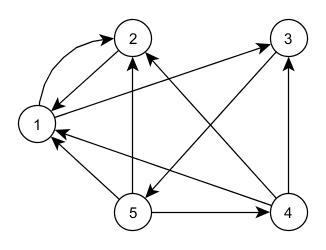}
\caption{Original graph}
\label{fig:figure1}
\end{minipage}
\hspace{0.5cm}
\begin{minipage}[b]{0.45\linewidth}
\centering
\includegraphics[width=\textwidth]{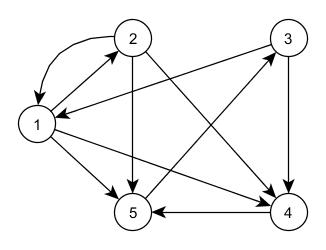}
\caption{Opposite graph}
\label{fig:figure2}
\end{minipage}
\end{figure}
\end{enumerate}
\section{No loops and multiple edges}
\label{31}
Algorithm removes multiple edges and loops from graph because they are irrelevant in process of finding Hamiltonian cycle.
\section{Unique neighbours}
\label{33}
Problem of finding Hamiltonian cycle in graph with $N$ vertices can be described as a problem of finding following edges:
$$a_{1} \rightarrow  x_{1}$$
$$a_{2} \rightarrow  x_{2}$$
$$\vdots$$
$$a_{N} \rightarrow  x_{N}$$
where ${x_{1}, x_{2}, \dots , x_{N} \in \{{a_{1}, a_{2}, \dots , a_{N}}}\}$ ;  $x_{1} \ne x_{2} \ne \dots \ne x_{N}$ and edges\\ $a_{1} \rightarrow  x_{1}, a_{2} \rightarrow  x_{2}, \dots , a_{N} \rightarrow  x_{N}$ don't create a cycle that is not Hamiltonian. \\ \newline
For graph with $N$ vertices algorithm should find $N$ unique neighbours.
\newline
\newline
Let's consider example graph presented with adjacency list: 
\newline $\textbf{1}: {\color{blue}3}, {\color{blue}4}$
\newline $\textbf{2}: {\color{blue}3}, {\color{blue}4}$
\newline $3: 2, 4$
\newline $4: 1, 3, 5$
\newline $5: 2, 3$
\newline $6: 1$
\newline
\newline
Let's test which neighbours have vertices $1$ and $2$. They are: $3$, $4$, $3$ and $4$.
\newline
Let's test how many and which unique neighbours are on the list.
\newline There are 2 vertices: $3$ and $4$.
\newline
\newline
\textbf{
When for $M$ tested vertices the number of unique neighbours equals $M$ then algorithm can remove these unique neighbours from vertices, that were not tested. $M$ is lower than $N$.} 
\newline \newline Graph after removal of unnecessary edges presented with adjacency list:
\newline $\textbf{1}: {\color{blue}3}, {\color{blue}4}$
\newline $\textbf{2}: {\color{blue}3}, {\color{blue}4}$
\newline $3: 2, {\color{red}\cancel{4}} $
\newline $4: 1, {\color{red}\cancel{3}}, 5$
\newline $5: 2, {\color{red}\cancel{3}}$
\newline $6: 1$
\newline
\newline
\textbf{''Unique neighbours'' test is used for original graph and opposite graph.}
\newline
\newline
Let's consider another example graph presented with adjacency list: 
\newline $1: 2, 4$
\newline $2: 1, 2$
\newline $\textbf{3}: {\color{blue}2}$
\newline $4: 1, 4, 5$
\newline $5: 2, 3$
\newline $6: 1, 4$
\newline
\newline
Let's test which neighbours vertex  $3$ has. It's only $2$.
\newline
Let's test how many and which unique neighbours are on the list. There is only one vertex: $2$.
\newline Graph after removal of unnecessary edges presented with adjacency list:
\newline $1: {\color{red}\cancel{2}}, 4$
\newline $2: 1, {\color{red}\cancel{2}}$
\newline $\textbf{3}: {\color{blue}2}$
\newline $4: 1, 4, 5$
\newline $5: {\color{red}\cancel{2}}, 3$
\newline $6: 1, 4$

\subsection{Method of accomplishing ''unique neighbours'' test}
\label{UN}
For graph $G$, for every vertex $V$ in $G$ algorithm takes its adjacency list and checks how many and which of others adjacency lists are subsets of $V$'s adjacency list.
\noindent Let's consider example graph presented with adjacency list:
\newline $1: 4, 8 \ \ \ \ \ \ \ \ \ \ 5: 1, 4, 8$
\newline $2: 1, 3, 5, 7 \ \ \ \ 6: 1, 3, 4, 5$
\newline $3: 4, 5, 6 \ \ \ \ \ \ \ 7: 3, 5$
\newline $4: 1, 3, 7 \ \ \ \ \ \ \ 8: 5, 7$\\

\noindent To analyze vertex $2$ with ''unique neighbours'' test algorithm takes its adjacency list  - $1, 3, 5, 7$ and checks how many and which of others adjacency lists are subsets of $2$'s adjacency list, they are:
\newline $4: 1, 3, 7 \ \ \ \ \ \ $
$7: 3, 5 \ \ \ \ $
$8: 5, 7$\\\\
Algorithm can remove edges marked with red color:
\newline $1: 4, 8 \ \ \ \ \ \ \ \ \ \  5: {\color{red}\cancel{1}}, 4, 8$
\newline ${\color{blue}2: 1, 3, 5, 7} \ \ \ \  6: {\color{red}\cancel{1}}, {\color{red}\cancel{3}}, 4, {\color{red}\cancel{5}}$
\newline $3: 4, {\color{red}\cancel{5}}, 6 \ \ \ \ \ \ \  {\color{blue}7: 3, 5}$
\newline {\color{blue}$4: 1, 3, 7 \ \ \ \ \ \ \  8: 5, 7$}
\subsection{Add edges to remove edges}
\label{add}
\noindent Let's consider example graph presented with adjacency list:
\newline $1: 5, 6, 7 \ \ \ \ \ \ 5: 2, 7$
\newline $2: 1, 4 \ \ \ \ \ \ \ \ \ 6: 3, 2, 5$
\newline $3: 2, 5, 7 \ \ \ \ \ \ 7: 1, 4, 5$
\newline $4: 2, 5, 6$\\

\noindent When algorithm tests graph with method \ref{UN} it will not find any edges that it could remove. However with addition of edge $1 \rightarrow 2$ algorithm can remove edges marked with red color:

\noindent
 {\color{blue}$1: \textbf{2},5, 6, 7 \ \ \ 5: 2, 7$}
\newline $2: 1, 4 \ \ \ \ \ \ \ \ \ 6: 3, {\color{red}\cancel{2}}, {\color{red}\cancel{5}}$
\newline ${\color{blue}3: 2, 5, 7} \ \ \ \ \ \  7: 1, 4, {\color{red}\cancel{5}}$
\newline {\color{blue}$4: 2, 5, 6$}\\
 Temporary addition of new edges may allow algorithm to remove some edges.
\subsubsection{Which edges should be added?}
\label{addMethod}
Let's consider graph with $N$ vertices:  ${x_{1}, x_{2}, \dots , x_{N}}$. To test which edges should be added while testing vertex $x_{1}$ with method \ref{UN} algorithm compares adjacency list of $x_{1}$ with adjacency lists of ${x_{2}, x_{3}, \dots , x_{N}}$:\\
If lists $x_{1}$ and $x_{2}$ have at least one common vertex then added edges are these that are on the list $x_{2}$ but that are not on list $x_{1}$.\\
If lists $x_{1}$ and $x_{3}$ have \dots \\
\vdots \\
If lists $x_{1}$ and $x_{N}$ have \dots \\
\\ Let's consider example graph presented with adjacency list:
 \newline$ 1: 2, 5, 6, 7 \ \ \ \  5: 2, 4, 7, 8$
  \newline$ 2: 3, 5, 7 \ \ \ \ \ \ \ 6: 3, 4, 5, 7, 8$
  \newline$ 3: 2, 5, 7 \ \ \ \ \ \ \  7: 2, 5, 8$
  \newline$ 4: 1, 2, 7, 8 \ \ \ \  8: 1, 3, 5$\\
  \\Added edges(edges marked with red color are not tested):
\begin{table}[h]
\centering
\begin{tabular}{|c|c|c|c|c|c|c|c|c|}
\hline
- & $1$ & $2$               & $3$                                                                         & $4$               & $5$                                                                         & $6$ & $7$               & $8$                                                                         \\ \hline
$1$ & - & $1 \rightarrow 3$ &                                                                           & $1 \rightarrow 8$ & \begin{tabular}[c]{@{}c@{}}$1 \rightarrow 4$\\ $1 \rightarrow 8$\end{tabular} &   & {\color{red}$1 \rightarrow 8$} & {\color{red}$1 \rightarrow 3$}                                                           \\ \hline
$2$ &   & -               &                                                                           &                 &                                                                           &   & $2 \rightarrow $8 & $2 \rightarrow 1$                                                           \\ \hline
$3$ &   &                 & -                                                                         &                 &                                                                           &   & $3 \rightarrow 8$ & $3 \rightarrow 1$                                                           \\ \hline
$4$ &   &                 &                                                                           & -               &                                                                           &   & $4 \rightarrow 5$ & \begin{tabular}[c]{@{}c@{}}$4 \rightarrow 3$\\ $4 \rightarrow 5$\end{tabular} \\ \hline
$5$ &   &                 &                                                                           &                 & -                                                                         &   &                 &                                                                           \\ \hline
$6$ &   & $6 \rightarrow 1$ & \begin{tabular}[c]{@{}c@{}}$6 \rightarrow 2$\\ $6 \rightarrow 1$\end{tabular} &                 & $6 \rightarrow 2$                                                           & - & {\color{red}$6 \rightarrow 2$} & {\color{red}$6 \rightarrow 1$}                                                          \\ \hline
$7$ &   &                 &                                                                           &                 &                                                                           &   & -               & \begin{tabular}[c]{@{}c@{}}$7 \rightarrow 1$\\ $7 \rightarrow 3$\end{tabular} \\ \hline
$8$ &   &                 &                                                                           &                 &                                                                           &   &                 & -                                                                         \\ \hline
\end{tabular}
\end{table}
\newpage
\subsubsection{Combination of added edges}
\label{combination}
\noindent Let's consider example graph presented with adjacency list:
\newline $ 1: 4, 5, 7, 8, 9$ \ \ \ \ \ \ \ $ 6: 2, 7, 9, 10, 13, 14, 15$ \ \ \ \ \ \ \ \ \ \ \ \  $11: 12, 14$
\newline $ 2: 6, 9, 13, 14$ \ \ \ \ \ \ \ $ 7: 1, 6, 12, 13, 14$ \ \ \ \ \ \ \ \ \ \ \ \ \ \ \ \ \ \ \ $12: 5, 7, 8, 11$
\newline $ 3: 5, 8, 9, 15$ \ \ \ \ \ \ \ \  $ 8: 1, 3, 9, 12, 13$ \ \ \ \ \ \ \ \ \ \ \ \ \ \ \ \ \ \ \ \ \ $13: 2, 4, 6, 7, 8, 10, 15$
\newline $ 4: 1, 13, 14$ \ \ \ \ \ \ \ \ \ $ 9: 1, 2, 3, 6, 8$ \ \ \ \ \ \ \ \ \ \ \ \ \ \ \ \ \ \ \ \ \  \ \ \ $14: 2, 4, 6, 7, 10, 11$
\newline $ 5: 1, 3, 12, 15$ \ \ \ \ \ $10: 6, 13, 14$ \ \ \ \ \ \ \ \ \ \ \ \ \ \ \ \ \ \ \  \ \ \ \ \ \ \ $15: 3, 5, 6, 13$
\newline \newline Algorithm can temporary add a combination of added edges. Algorithm can temporary add following set of edges:
$11 \rightarrow 1,11 \rightarrow 3,11 \rightarrow 6,11 \rightarrow 9,11 \rightarrow 13$ which allows algorithm to remove edges marked with red color:
\newline $ 1: 4, 5, 7, 8, \color{red}{\cancel{9}}$ \ \ \ \ \ \ \ \ $ 6: 2, 7, \color{red}{\cancel{9}}\color{black}{, 10, }\color{red}{\cancel{13}}, \color{red}{\cancel{14}}\color{black}{, 15}$ \ \ \ \ \ \ \ \ \ \ \ \  $\color{blue}{\textbf{11}: \textbf{1}, \textbf{3}, \textbf{6}, \textbf{9}, 12, \textbf{13}, 14}$
\newline $ \color{blue}{2: 6, 9, 13, 14}$ \ \ \ \ \ \ \ \  $ \color{blue}{7: 1, 6, 12, 13, 14}$ \ \  \ \ \ \ \ \ \ \ \ \ \ \ \ \ \ \ \ \ $12: 5, 7, 8, 11$
\newline $ 3: 5, 8, \color{red}{\cancel{9}}\color{black}{, 15}$ \ \ \ \ \ \ \ \ \  $ \color{blue}{8: 1, 3, 9, 12, 13}$ \ \ \ \ \ \ \ \ \ \ \ \ \ \ \ \ \ \ \ \ \ \ $13: 2, 4, \color{red}{\cancel{6}}\color{black}{, 7, 8, 10, 15}$
\newline $ 4: \color{blue}{1, 13, 14}$ \ \ \ \ \ \ \ \ \ \ \ $ 9: \color{red}{\cancel{1}}\color{black}{, 2,} \color{red}{\cancel{3}}, \color{red}{\cancel{6}}\color{black}{, 8}$ \ \ \ \ \ \ \ \ \ \ \ \ \ \ \ \ \ \ \ \ \ \ \ \ \  $14: 2, 4, \color{red}{\cancel{6}}\color{black}{, 7, 10, 11}$
\newline $ 5: \color{red}{\cancel{1}}, \color{red}{\cancel{3}}, \color{red}{\cancel{12}}, \color{black}{15}$ \ \ \ \ \ \ $\color{blue}{10: 6, 13, 14}$ \ \ \ \ \ \ \ \ \ \ \ \ \ \ \ \ \ \ \ \ \ \ \ \ \ \ \  $\color{blue}{15: 3, 5, 6, 13}$
\\\\
\textbf{How to select combination of added edges:}

Algorithm tries to select added edges  in "greedy" way. It tries to reduce diffrence beetween:
\\1)  \textbf{$A$}'s degree with addition of number of added edges
\\2) amount of vertices which adjacency list is subset of \textbf{$A$}'s adjacency list with added edges
\\to 0 - to remove edges and below 0 - to determine that graph is not Hamiltonian.\\\\All possible subsets that can be added to $\textbf{11}$ in example graph are:
\\4, 5, 7, 8, 9  for 1
\\\textbf{6, 9, 13}  for 2
\\5, 8, 9, 15  for 3
\\\textbf{1, 13}  for 4
\\1, 3, 15  for 5
\\2, 7, 9, 10, 13, 16 for 6
\\1, 6, 13  for 7
\\\textbf{1, 3, 9, 13} for 8
\\1, 2, 3, 6, 8 for 9
\\6, 13  for 10
\\5, 7, 8, 11  for 12
\\2, 4, 6, 7, 8, 10, 15 for 13
\\2, 4, 6, 7, 10, 11 for 14
\\3, 5, 6, 13 for 15
\\\\Out of these subsets algorithm selects following combination: \textbf{1, 3, 6, 9, 13}.
\subsection{Not enough unique neighbours}
\label{NotEnoughUN}
Graph is not Hamiltonian when number of unique neighbours checked for any $M$ vertices is lower than $M$.\\\\ Example presented with adjacency list:
\newline $\color{red}{1: 4, 5}$
\newline $\color{red}{2: 4, 5}$
\newline $\color{red}{3: 4, 5}$
\newline $4: 1, 2$
\newline $5: 1, 3$
\\\\ Another example presented with adjacency list:
\newline $1: 1, 4, 5$
\newline $\color{red}{2: }$
\newline $3: 1, 4$
\newline $4: 2, 3$
\newline $5: 1, 2$
\subsection{Looped neighbours}
\label{LN}
Graph is not Hamiltonian when:
\newline 
1) for M tested vertices number of unique neighbours equals M 
\newline 2) subsets of: a) M tested vertices and b) theirs unique neighbours, are equal.
\newline
because in such a case, each possible assignment of edges in M vertices, forms a cycle that is not a Hamiltonian cycle. 
\newline
\newline
Example graph presented with adjacency list: 
\newline $\color{red}{\textbf{1}: 5, 6}$
\newline $2: 3, 4$
\newline $3: 6$
\newline $4: 2, 3$
\newline $\color{red}{\textbf{5}: 1, 6}$
\newline $\color{red}{\textbf{6}: 1, 5}$
\subsubsection{Almost looped neighbours}
\label{ALN}
When:
\newline 
1) for M tested vertices number of unique neighbours equals M + 1 and
\newline 2) M tested vertices contains only one unique neighbour that was not in M tested vertices
\newline
then this one unique neighbour that was not in M tested vertices, must be in Hamiltonian cycle, if Hamiltonian cycle exists in graph. Therefore algorithm can delete edges other than this one unique neighbor. 
\newline
\newline
Example graph presented with adjacency list: 
 \newline $1: 4, 5, 6, 7$
 \newline $2: 3, 4$
 \newline $3: 1$
 \newline $4: 1, 2$
 \newline $\textbf{5}: \color{OliveGreen}{4}, \color{red}{\cancel{6}, \cancel{7}}$
 \newline $\textbf{6}: \color{blue}{5, 7}$
 \newline $\textbf{7}: \color{blue}{5, 6}$
\subsection{"Looped neighbours" effect}
\label{LNE}
To analyze vertex $x$ with ''unique neighbours'' test algorithm takes its adjacency list and checks how many and which of others adjacency lists are subsets of $x$'s adjacency list. Algorithm also checks how many and which of others adjacency lists \textbf{can be} subsets of $x$'s adjacency list - rest of it's neighbours are not on adjacency list of $x$. Then algorithm only takes under consideration those neighbours which are not on adjacency list of $x$ - if they create effect of "looped neighbours", than one of those vertices will not be in Hamiltonian cycle, meaning this vertex will definetly be a subset of $x$'s adjacency list. Because of that algorithm can remove edges or determine that graph is not Hamiltonian.
\\
\\
Example 1 - graph presented with adjacency list: 
\newline $1: \color{red}{\cancel{2}}\color{black}{, 3, 7}$
\newline $2: 1, 3, \color{red}{\cancel{8}}$
\newline $ 		\textbf{3: \color{blue}{2, 5, 6, 8}}$
\newline $     \color{Orange}{4}: \color{blue}{6}, \color{YellowOrange}{7}$
\newline $ 	\textbf{5}: \color{blue}{6, 8}$
\newline $ 	\textbf{6}: \color{blue}{2, 5, 8}$
\newline $     \color{Orange}{7}: \color{blue}{2}, \color{YellowOrange}{4}, \color{blue}{5}$
\newline $ 8: 1, 3, \color{red}{\cancel{6}}$
\newline To analyze vertex $\textbf{3}$ with ''unique neighbours'' test algorithm takes its adjacency list and checks how many and which of others adjacency lists are subsets of $\textbf{3}$'s adjacency list. They are:  $\textbf{5}$ and $\textbf{6}$. Then algorithm checks how many and which of others adjacency lists are \textbf{can be} subsets of $\textbf{3}$'s adjacency list - they are $\textbf{4}$ and $\textbf{7}$. Than algorithm checks if $\textbf{4}$ and $\textbf{7}$ create an effect of "looped neighbours" - it is true. In summary, there are following subsets of  $\textbf{3}$'s adjacency list:
\newline 1. $\textbf{3}$'s adjacency list.
\newline 2. $\textbf{5}$'s adjacency list.
\newline 3. $\textbf{6}$'s adjacency list.
\newline 4. $\textbf{4}$'s or $\textbf{7}$'s adjacency list.
\newline therefore algorithm can remove edges: $\textbf{1}$ $\rightarrow$ $\textbf{2}$, $\textbf{2}$ $\rightarrow$  $\textbf{8}$, $\textbf{8}$ $\rightarrow$  $\textbf{6}$.
\\
\\
Example 2 - graph presented with adjacency list: 
\newline $ 1: 2, 3, 5, 6, 9, \color{red}{\cancel{12}}, \color{red}{\cancel{15}}$
\newline $ \color{Orange}{2}: \color{blue}{1}, \color{YellowOrange}{5}, \color{YellowOrange}{7}$
\newline $ \color{Orange}{3}: \color{blue}{1}, \color{YellowOrange}{10}$
\newline $ 4: 6, 10, 13$
\newline $ \color{Orange}{5}: \color{blue}{1}, \color{YellowOrange}{2}, \color{YellowOrange}{7}$
\newline $ \textbf{6: \color{blue}{1, 4, 12, 14, 15}}$
\newline $ \color{Orange}{7}: 2, 5, \color{blue}{12}$
\newline $ \color{Orange}{8}: 10, \color{blue}{14}$
\newline $ \textbf{9}: \color{blue}{1, 15}$
\newline $ \color{Orange}{10}: \color{YellowOrange}{3}, \color{blue}{4}, \color{YellowOrange}{8}, \color{YellowOrange}{11}, \color{blue}{12}$
\newline $\color{Orange}{11}: \color{YellowOrange}{10}, \color{blue}{15}$
\newline $12: \color{red}{\cancel{1}}\color{black}{, 6, 7, 10, 13}$
\newline $\textbf{13}: \color{blue}{4, 12, 15}$
\newline $14: 6, 8$
\newline $15: \color{red}{\cancel{1}}\color{black}{, 6, 9, 11, 13}$
\\There are following subsets of  $\textbf{6}$'s adjacency list:
\newline 1. $\textbf{6}$'s adjacency list.
\newline 2. $\textbf{9}$'s adjacency list.
\newline 3. $\textbf{13}$'s adjacency list.
\newline 4. $\textbf{2}$'s or $\textbf{5}$'s or $\textbf{7}$'s adjacency list.
\newline 5. $\textbf{3}$'s or $\textbf{8}$'s or $\textbf{10}$'s or $\textbf{11}$'s adjacency list.
\newline therefore algorithm can remove edges: $\textbf{1}$ $\rightarrow$ $\textbf{12}$, $\textbf{1}$ $\rightarrow$  $\textbf{15}$, $\textbf{12}$ $\rightarrow$  $\textbf{1}$, $\textbf{15}$ $\rightarrow$  $\textbf{1}$.
\subsection{Recursion effect - "unique neighbours among unique neighbours"}
\label{recursion}
In addition to the above-described method of finding which ajdacency lists can be subsets of $x$'s adjacency list, algorithm also recursively uses the "unique neighbors" principle.\\ Let's consider following graph presented with adjacency list: 
\newline $\textbf{1: \color{blue}{2, 3, 5, 7}}$
\newline $\textbf{2}: \color{blue}{5, 7}$
\newline $\color{Orange}{3}: \color{YellowOrange}{4}, \color{blue}{5}$
\newline $\color{Orange}{4}: \color{YellowOrange}{3}, \color{blue}{5}$
\newline $\color{Green}{5}: \color{blue}{2}, \color{blue}{3}, \color{black}{8}$
\newline $\color{Green}{6}: \color{blue}{5}, \color{blue}{7}, \color{black}{8}$
\newline $7: 1, \color{red}{\cancel{4}}, \color{red}{\cancel{5}}$
\newline $8: \color{red}{\cancel{4}}, \color{red}{\cancel{5}}$
\\
\\
Thanks to "unique neighbors" principle, algorithm knows that if Hamiltonian cycle exists in graph, only one of following statements is true:
\begin{itemize}
\item Hamiltonian cycle contains edge $\textbf{5}$ $\rightarrow$ $\textbf{8}$
\item Hamiltonian cycle contains edge $\textbf{6}$ $\rightarrow$ $\textbf{8}$
\item Hamiltonian cycle doesn't contain both $\textbf{5}$ $\rightarrow$ $\textbf{8}$ and $\textbf{6}$ $\rightarrow$ $\textbf{8}$ edges
\end{itemize}

That's why, there are following subsets of  $\textbf{1}$'s adjacency list:
\newline 1. $\textbf{1}$'s adjacency list.
\newline 2. $\textbf{2}$'s adjacency list.
\newline 3. $\textbf{3}$'s or $\textbf{4}$'s adjacency list, because of "looped neighoburs" effect.
\newline 2. $\textbf{5}$'s or $\textbf{6}$'s adjacency list, because of "unique neighbours among unique neighbours" effect.
\\Therefore algorithm can remove edges: $\textbf{7}$ $\rightarrow$ $\textbf{4}$, $\textbf{7}$ $\rightarrow$  $\textbf{5}$, $\textbf{8}$ $\rightarrow$  $\textbf{4}$, $\textbf{8}$ $\rightarrow$  $\textbf{5}$.

\section{Single edge in only one direction}
\label{SingleEdge}
When in original graph or in opposite graph the only one neighbour of vertex $X$ is vertex $Y$, it means that for Hamiltonian cycle to exist edge $X \rightarrow Y$ must be in this cycle, when graph also contains edge $Y \rightarrow X$ then algorithm can remove edge $Y \rightarrow X$.
\\
\\
Example graph presented with adjacency list: 
\newline $1: 2, 3, 4$
\newline $\textbf{2: 3}$
\newline $\textbf{3:  \color{red}{\cancel{2}}} \color{black},4$
\newline $4: 1$
\\
\\
Algorithm can remove edge $3 \rightarrow 2$.
\subsection{Path that must be in Hamiltonian cycle}
\label{SinglePath}
When in original graph or in opposite graph:
\begin{itemize}
\item the only one neighbour of vertex $A$ is vertex $B$, when graph also contains edge $B \rightarrow A$ then algorithm can remove edge $B \rightarrow A$
\item when also the only one neighbour of vertex $B$ is vertex $C$, when graph also contains edge $C \rightarrow A$ then algorithm can remove edge $C \rightarrow A$
\newline $ \vdots $
\item when also the only one neighbour of vertex $Y$ is vertex $Z$, when graph also contains edge $Z \rightarrow A$ then algorithm can remove edge $Z \rightarrow A$
\end{itemize}
In summary, if path $A \rightarrow B \rightarrow C \dots \rightarrow Z$ must be in Hamiltonian cycle than algorithm can remove edges:
\begin{itemize}
\item $B \rightarrow A$, if it exists in graph 
\item $C \rightarrow A$, if it exists in graph 
\newline $ \vdots $
\item $Z \rightarrow A$, if it exists in graph 
\end{itemize}
\noindent
\newline
Example 1 - graph presented with adjacency list: 
\newline $1: 5$
\newline $2: 6, \color{red}{\cancel{7}}$
\newline $\color{blue}{\textbf{3}: 2}$
\newline $4: 1$
\newline $5: 3, 4, 6$
\newline $6: 1, 4, 7$
\newline $\color{blue}{\textbf{7}: 3}$
\newline If graph contains Hamiltonian cycle, than path $7 \rightarrow 3 \rightarrow 2$ must be in it, therefore algorithm can remove edge $2 \rightarrow 7$.
\newline
\newline
Example 2 - graph presented with adjacency list: 
\newline $\color{blue}{\textbf{1}: 2}$
\newline $2: \color{red}{\cancel{4}}, \color{black}{6}$
\newline $3: \color{red}{\cancel{4}}, \color{black}{5, 7}$
\newline $\color{blue}{\textbf{3}: 1}$
\newline $\color{blue}{\textbf{4}: 3}$
\newline $5: 4, 6, 7$
\newline $6: 4, 5, 7$
\newline $7: 4, 5, 6$
\newline If graph contains Hamiltonian cycle, than path $4 \rightarrow 3 \rightarrow 1 \rightarrow 2$ must be in it, therefore algorithm can remove edges $3 \rightarrow 4$ and $2 \rightarrow 4$.
\section{Cycle that is not Hamiltonian}
\label{cykl}
Graph is not Hamiltonian when: 
\subsection{Vertices with 1 neighbour}
\label{Vertices1}
In original graph or in opposite graph among vertices that have 1 neighbour exists cycle, that is not Hamiltonian. Example graph:
\newline
\begin{figure}[h!]
    \centering
    \includegraphics[width=0.5\textwidth]{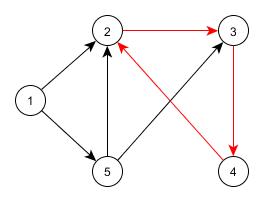}
    \caption{Cycle created by vertices: $2$, $3$ and $4$}
\end{figure}
\subsection{Vertices with more than 1 neighbour}
\label{VerticesMoreThan1}
Let's consider graph with $N$ vertices. Graph contains vertex X that has $M$ neighbours, $1 < M < N / 2 $, $X$'s list of in-neighbours is a superset of $X$'s list of out-neighbours. Graph also contains at least $M-1$ vertices which their list of in-neighbours are supersets of out-neighbours and list of in-neighbours which is a subset of list of neighbours of vertex $X$.\newline \newline \noindent Example 1. presented with adjacency list:\\
Original and opposite graph:\\
X: A B\\
Y: A B\\
\\There are only 2 paths that can be created, they are:
\begin{itemize}
\item $A \rightarrow X \rightarrow B \rightarrow Y \rightarrow A$ 
\item $B \rightarrow X \rightarrow A \rightarrow Y \rightarrow B$ 
\end{itemize}
Every possible path creates cycle that is not Hamiltonian.\\\\
Example 2. presented with adjacency list:\\
Original and opposite graph:\\
X: A B C\\
Y: A B C\\
Z: A B C\\
\\There are 12 paths that can be created, they are:
\begin{itemize}
\item $A \rightarrow X \rightarrow B \rightarrow Y \rightarrow C \rightarrow Z \rightarrow A$ 
\item $A \rightarrow X \rightarrow B \rightarrow Z \rightarrow C \rightarrow Y \rightarrow A$ 
\item $A \rightarrow X \rightarrow C \rightarrow Y \rightarrow B \rightarrow Z \rightarrow A$ 
\item $A \rightarrow X \rightarrow C \rightarrow Z \rightarrow B \rightarrow Y \rightarrow A$ 
\item $B \rightarrow X \rightarrow A \rightarrow Y \rightarrow C \rightarrow Z \rightarrow B$ 
\item $B \rightarrow X \rightarrow A \rightarrow Z \rightarrow C \rightarrow Y \rightarrow B$ 
\item $B \rightarrow X \rightarrow C \rightarrow Y \rightarrow A \rightarrow Z \rightarrow B$ 
\item $B \rightarrow X \rightarrow C \rightarrow Z \rightarrow A \rightarrow Y \rightarrow B$ 
\item $C \rightarrow X \rightarrow A \rightarrow Y \rightarrow B \rightarrow Z \rightarrow C$ 
\item $C \rightarrow X \rightarrow A \rightarrow Z \rightarrow B \rightarrow Y \rightarrow C$ 
\item $C \rightarrow X \rightarrow B \rightarrow Y \rightarrow A \rightarrow Z \rightarrow C$ 
\item $C \rightarrow X \rightarrow B \rightarrow Z \rightarrow A \rightarrow Y \rightarrow C$ 
\end{itemize}
Every possible path creates cycle that is not Hamiltonian.
\noindent
\newline
\newline
Example 3. presented with adjacency list:\\
Original and oppposite graph:\\
X: A B C\\
Y: A B\\
Z: A C\\
\\There are 2 paths that can be created, they are:
\begin{itemize}
\item $B \rightarrow X \rightarrow C \rightarrow Z \rightarrow A \rightarrow Y \rightarrow B$ 
\item $C \rightarrow X \rightarrow B \rightarrow Y \rightarrow A \rightarrow Z \rightarrow C$ 
\end{itemize}
Every possible path creates cycle that is not Hamiltonian.
\section{Closed alleys}
\label{Alley}
If in the graph any of the vertices has a neighbour, the selection of which creates a path that is not a Hamiltonian cycle, and which creates a path that ends back at the same vertex, the edge with that neighbour can be deleted. 
\newline
\newline
Example  presented with adjacency list:
 \newline $\color{blue}{\textbf{1}: 7}$
 \newline $2: 4, 5, 6$
 \newline $3: 2$
 \newline $\color{blue}{\textbf{4}: 1}$
 \newline $5: 3, 4, 6$
 \newline $6: 3, 4$
 \newline $\color{blue}{\textbf{7}}: \color{black}{3, } \color{red}{\textbf{\cancel{4}}}\color{black}{, 5, 6}$
\newline
If edge $ 7 \rightarrow 4 $ would be in Hamiltonian cycle, it would create a path: $7 \rightarrow 4 \rightarrow 1 \rightarrow 7$, such a path would create a cycle that is not Hamiltonian, therefore algorithm can remove edge $ 7 \rightarrow 4 $.
\section{Graph is disconnected}
\label{GraphDisconnected}
Let's consider graph with $N$ vertices: $a_1, a_2, \dots , a_N$.If a path doesn't exist between vertices: $a_1$ to $a_2$ or $a_1$ to $a_3$ \dots or $a_1$ to $a_N$, then graph is not Hamiltonian.
\begin{figure}[h!]
    \centering
    \includegraphics[width=0.95\textwidth]{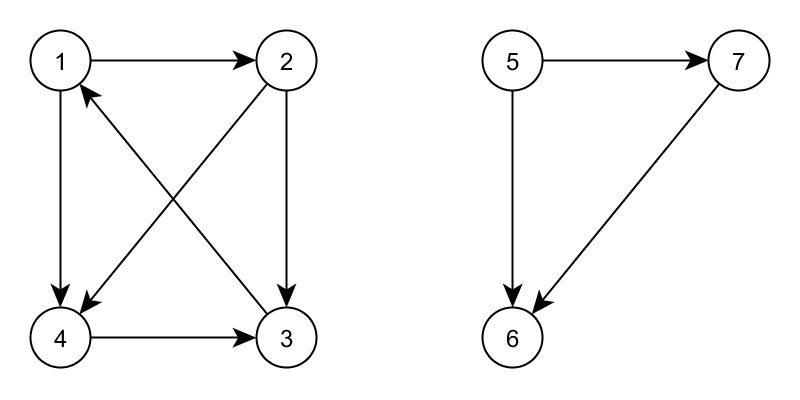}
    \caption{A path doesn't exist beetween vertices $1$ and $5$}
\end{figure}
\section{Never only 1 Hamiltonian cycle}
In \textbf{undirected} graph for each edge $ a \rightarrow  b $
there is an edge
$ b \rightarrow  a $
, which means that if there is a Hamiltonian cycle in an undirected graph: $(v_1,v_2)$, $(v_2,v_3)$, \dots , $(v_{K-1},v_K)$ the following Hamiltonian cycle must also exist in this graph: $(v_K,v_{k-1})$, \dots , $(v_3,v_2)$ $(v_2,v_1)$. \textbf{This means that if in an undirected graph, which contains Hamiltonian cycle, there is a vertex that has exactly 2 degree, then both edges coming from this vertex must be in  Hamiltonian cycle.}
\subsection{Too low degree}
\label{Never1HC_TLD}
\textbf{Graph is not Hamiltonian} when in an undirected graph there is any vertex that has degree lower than 2.  \newline Example - graph presented with adjacency list: 
 \newline $0: 2, 3$
 \newline $1: \color{red}{2}$
 \newline $2: 0, 1, 4$
 \newline $3: 0, 4$
\newline $ 4: 2, 3$
\subsection{Too many similar edges}
\label{Never1HC_SE}
If among above-described necessary edges for the existence of a Hamiltonian cycles, there are more than two edges with the same out-neigbour, it means that \textbf{graph is not Hamiltonian}.
\newline Example - graph presented with adjacency list: 
 \newline $ 0: \color{red}{\textbf{5}, 6}$
  \newline $1: 3, 4, 6$
 \newline $ 2: \color{red}{4, \textbf{5}}$
 \newline $ 3: \color{red}{1, \textbf{5}}$
  \newline $4: 1, 2, 6$
  \newline $5: 0, 2, 3$
  \newline $6: 0, 1, 4$
\newline It is impossible for Hamiltonian cycles to contain all of the following edges simultaneously: $ 0 \rightarrow 5 $, $ 2\rightarrow 5 $ and $ 3 \rightarrow 5 $.
\subsection{Remove edges}
\label{Never1HC_RE}
If graph has a Hamiltonian cycle and graph has a vertex $x$, which has exactly 2 degree, $x: a, b$, it means that the graphs have at least two Hamiltonian cycles containing edges: $ x \rightarrow a $, $ x \rightarrow b $, $ a \rightarrow x $, $ b \rightarrow x $. Based on this observation, algorithm can remove edges, assuming that the above-indicated edges must be in the Hamiltonian cycle.
\newline Example - graph presented with adjacency list: 
   \newline $0: \color{red}{\cancel{4}} \color{black}{, 5, 6}$
   \newline $1: 3, 4, 5 $
   \newline $2: 3, 4, 6$
   \newline $3: 1, 2, 4 $
   \newline $4: \color{red}{\cancel{0}}\color{black}{, 1, 2, 3} $
   \newline $5: \color{blue}{\textbf{0}, 1} $
   \newline $6: \color{blue}{\textbf{0}, 2} $
   \newline If above graph has Hamitlonian cycles, then these cycles contain edges $ 5 \rightarrow 0 $, $ 5 \rightarrow 1 $, $ 6 \rightarrow 0 $ and $ 6 \rightarrow 1 $.
  Since these cycles contain edges: $ 5 \rightarrow 0 $ and $ 6 \rightarrow 0 $ , then edge $ 4 \rightarrow 0 $ can be deleted from the graph, and due to fact that the graph is undirected, then edge $ 0 \rightarrow 4 $ can also be deleted.
\subsection{Edges constantly removed}
\label{Never1HC_FECR}
If undirected graph has Hamiltonian cycle, and this graph has a vertex $x$ with a degree greater than $2$ - $x: a, b, \dots z$ where \textbf{only} $a$ has a degree $2$, it means that possibly existing Hamiltonian cycles in this graph, will definitely contains an edge $x \rightarrow a$ and one of the following edges:
\newline $x \rightarrow b$
\newline  \dots
\newline  $x \rightarrow z$
\newline \newline Example - graph presented with adjacency list: 
   \newline $ 0: 2, \color{blue}{3}$
    \newline $1: 4, 5$
    \newline $2: 0, 6$
    \newline $3: \color{blue}{0},\color{YellowOrange}{4, 6}$
    \newline $4: 1, 3, 5$
    \newline $5: 1, 4, 6$
    \newline $6: 2, 3, 5$
 \newline
If above graph has Hamiltonian cycle, it definitely contains an edge $0 \rightarrow 3$ and one of the following edges: $3 \rightarrow 4$  or $4 \rightarrow 6$ .
\newline  
 \textbf {Based on this observation, algorithm can remove edges, with methods described in \ref{FECR}}
\section{Edges constantly removed}
\label{FECR}
\subsection{Remove edge}
Algorithm performs following test for every edge in graph: if after removal of a single edge in graph analysis of graph proves that graph is not Hamiltonian, then for a Hamiltonian cycle to exist in graph, such an edge must be in it. 
\newline Therefore if vertex \textbf{$X$}'s adjacency list consists of $A,B,C, \dots Z$ and edge \textbf{$X \rightarrow A$} is a "neccesary edge" then algorithm can remove edges: \textbf{$X \rightarrow B$}, \textbf{$X \rightarrow C$}, \dots \textbf{$X \rightarrow Z$}.
\newline 
\newline Example  presented with adjacency list:
\newline  $1: 2, 4, 5, 6$
\newline  $\textbf{2}: \textbf{\color{blue}{3}}, \color{red}{\cancel{4}, \cancel{6}, \cancel{8}}$
\newline  $3: 4, 6, 7, 8$
\newline  $4: 1, 6, 8$
\newline  $5: 1, 2, 4, 6$
\newline  $6: 1, 4, 8$
\newline  $7: 1, 2$
\newline  $8: 4, 5, 6, 7$
\newline After removal of edge $2 \rightarrow 3$ analysis of graph proves that graph is not Hamiltonian, therefore algorithm can remove edges: \textbf{$2 \rightarrow 4$}, \textbf{$2 \rightarrow 6$}, \textbf{$2 \rightarrow 8$}.
\subsection{Select edge}
\label{ecr}
\begin{figure}[h!]
    \centering
    \includegraphics[width=0.56\textwidth]{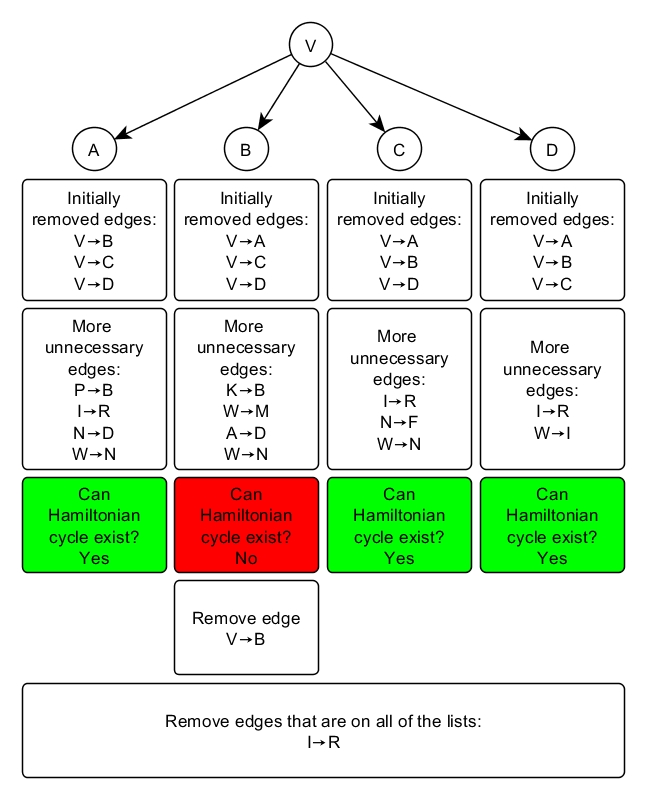}
    \caption{$I \rightarrow R$ is an edge constantly removed}
\end{figure}
\noindent
Graph with $N$ vertices: $a_1, a_2, \dots , a_N$. For Hamiltonian cycle to exist each of these vertices must contain an edge that is in this cycle. Vertex $V$ has $M$ neighbours: $ b_1, b_2, \dots , b_M; M\ge2; b_1, b_2, \dots , b_M\in\{a_1, a_2, \dots , a_N\}$ . After removal of certain edges, algorithm may be able to remove more unnecessary edges. Following tests are made for every vertex that has more than 1 neighbour, tests are based on an assumption that chosen edge must be in Hamiltonian cycle:
\begin{itemize}
\item Chosen edge: $V \rightarrow b_1$. Test which edges will be removed after removal of edges: $V \rightarrow b_2, V \rightarrow b_3, \dots ,  V \rightarrow b_M$, all removed edges are saved on list $l_1$
\item Chosen edge: $V \rightarrow b_2$. Test which edges will be removed after removal of edges: $V \rightarrow b_1, V \rightarrow b_3, \dots ,  V \rightarrow b_M$, all removed edges are saved on list $l_2$
\item \dots
\item Chosen edge: $V \rightarrow b_M$. Test which edges will be removed after removal of edges: $V \rightarrow b_1, V \rightarrow b_2, \dots ,  V \rightarrow b_{M-1}$, all removed edges are saved on list $l_M$
\end{itemize}
If test of following assumption: edge $V \rightarrow b_x$, where $x \in \lbrace 1,2, \dots , M \rbrace$, must be in Hamiltonian cycle, proves that Hamiltonian cycle can't exist if edge $V \rightarrow a_x$ is in it, then algorithm can remove edge $V \rightarrow b_x$ and doesn't save removed edges on list $l_x$. Example:  edge $V \rightarrow B$ shown in Figure 5.\\
When for all of the edges: $V \rightarrow b_1, V \rightarrow b_2, \dots , V \rightarrow b_M$ test will prove that Hamiltonian cycle can't exist if chosen edge is in it, then graph does not contain Hamiltonian cycle.\\
Edges that are on all of the lists: $l_1, l_2, l_3, \dots , l_M$ can be removed from graph because for Hamiltonian cycle to exist at least one of the tested edges must be chosen and regardless which edge will it be, the edges that are on all of the lists will be constantly removed. Example:  edge $I \rightarrow R$ shown in Figure 5.
\subsection{Compare removed edges}
For each edge in the graph, the following statement is true: if there is a Hamiltonian cycle in the graph, then the edge will either be in it or not. So the algorithm checks for each edge $A \rightarrow B$: if there are any edges that will be deleted, regardless of whether the edge $A \rightarrow B$ will be in the Hamiltonian cycle or not, then these edges can be deleted from the graph.
\section{Connected edges}
\label{CE}
If in \textbf{undirected} graph vertex $x$ has 2 degree - $X: A, B$ and graph has a Hamiltonian cycle, it means that: 
\begin{enumerate}
\item Edges $X \rightarrow A$ and $X \rightarrow B$ are parts of Hamiltonian cycles 
\item Edge $X \rightarrow A$ and edge $X \rightarrow B$ are not part of the same Hamiltonian cycle - they are in separate Hamiltonian cycles 
\end{enumerate}
In any undirected graph having a Hamiltonian cycle, there are at least 2 Hamiltonian cycles. In order to conclude that a graph has a Hamiltonian cycle, it is enough to find only 1 Hamiltonian cycle, it is not necessary to find a second Hamiltonian cycle, nor all of the Hamiltonian cycles in graph. 
\newline
\newline Edges can be "connected" by establishing which edges must exist simultaneously in the same Hamiltonian cycle. The attempt to connect the edges only occurs between the edges coming out of the vertices of degree 2. Let's consider following graph: 
\newline $A: X_1, X_2$
\newline $B: X_3, X_4$
\newline \dots
\newline $C: X_{N-1}, X_N$
\newline $Y: X_1, X_2, X_3, \dots$
\newline $Z: : X_1, X_2, X_3,\dots$
\newline \dots
\newline
\newline
Algorithm will try to connect edges: $A \rightarrow X_1$, $A \rightarrow X_2$, $B \rightarrow X_3$, $B \rightarrow X_4$, \dots ,  $C \rightarrow X_{N-1}$ and  $C \rightarrow X_{N}$
\newline
\newline
Consider the edges $A \rightarrow X_1$, $A \rightarrow X_2$, $B \rightarrow X_3$ and $B \rightarrow X_4$:
\begin{enumerate}
\item If, as a result of analysis, algorithm finds that $A \rightarrow X_1$ and $B \rightarrow X_3$ edges cannot exist simultaneously in the same Hamiltonian cycle, it means that the $A \rightarrow X_1$ and  $B \rightarrow X_4$ edges can be connected.
\item If \dots $A \rightarrow X_1$ and $B \rightarrow X_4$ \dots it means \dots $A \rightarrow X_1$ and  $B \rightarrow X_3$ edges can be connected.
\item If \dots $A \rightarrow X_2$ and $B \rightarrow X_3$ \dots it means \dots $A \rightarrow X_2$ and  $B \rightarrow X_4$ edges can be connected.
\item If \dots $A \rightarrow X_2$ and $B \rightarrow X_4$ \dots it means \dots $A \rightarrow X_2$ and  $B \rightarrow X_3$ edges can be connected.
\end{enumerate}
Algorithm tries to connect as many edges as possible, knowing that if  graph has a Hamltonian cycle, it would connect edges that are in Hamiltonian cycle. 
\newline \newline Example of connected edges:
\newline  $0: 3, 4, 5$
\newline  $ 1: \textbf{\color{blue}{2}}, \color{red}{\cancel{4}}$
\newline  $ 2: \color{red}{\cancel{1}}, \textbf{\color{blue}{6}}$
\newline  $ 3: 0, 4, 5$
\newline  $ 4: 0, 1, 3, 5, 6$
\newline  $5: 0, 3, 4, 7$
\newline  $6: 2, 4, 7$
\newline  $ 7: \textbf{\color{blue}{5}}, \color{red}{\cancel{6}}$
\section{Skipping search}
\label{DiracsTheorem}
For some graphs, algorithm is able to determine that graph is Hamiltonian based on:
\subsection{Bondy-Chvátal theorem}
\label{closure}
''The Bondy–Chvátal theorem operates on the closure cl(G) of a graph G with n vertices, obtained by repeatedly adding a new edge uv connecting a nonadjacent pair of vertices u and v with deg(v) + deg(u) $\ge$ n until no more pairs with this property can be found. "\\
\textbf{Bondy–Chvátal theorem}
    \par "A graph is Hamiltonian if and only if its closure is Hamiltonian.''\cite{Wikipedia}.
\subsection{Dirac's theorem}
 "A simple graph with n vertices (n $\ge$3) is Hamiltonian if every vertex has degree $\frac{n}{2}$ or greater."\cite{Wikipedia}.
\subsection{Adding edges}
Let's consider graph G with M edges. If G is Hamiltonian, it will always remain Hamiltonian, regardless of how many edges could be added to G.
\subsection{Directed graphs}
Bondy-Chvátal theorem and Dirac's theorem apply to undirected graphs. These theorems can be applied to directed graphs, but only it those graphs are changed into their undirected verions, by removing all edges $A \rightarrow B$, if graph doesn't contain edge $B \rightarrow A$.   
\subsection{Summary}
Algorithm uses Dirac's theorem on closure of:
\begin{enumerate}
\item undirected graphs - if graph is "fully undirected"
\item "undirected version" of directed graph - if tested graph is not "fully undirected"
\end{enumerate}

\section{Most optimal path}
\label{5}
\subsection{Visiting a path}
Both brute-force search and algorithm use recursive depth-first search to test paths - test possibilities, however the second one use it differently. Algorithm's goal is to find edges described in the beginning of \ref{33}.\\ \\ Let's consider graph $G$ with $N$ vertices and path $P$ from graph $G$ which consists of following edges:
$a_{1} \rightarrow  a_{2}, a_{2} \rightarrow  a_{3}, \dots , a_{M-1} \rightarrow  a_{M}$.\\ \\
\textbf{Brute-force search} will visit $P$ in following way:\\
1. Visit vertex $a_{1}$\\
2. Visit vertex $a_{2}$\\
\vdots \\
M. Visit vertex $a_{M}$\\
M+1. Check if $N==M$ and if $G$ contains edge $a_{M} \rightarrow  a_{1}$
\\
\\
\textbf{Alghorithm} will visit $P$ in following way:\\
1. Visit edge $a_{x_1} \rightarrow  a_{x_2}$\\
2. Remove unnecessary edges and test if Hamiltonian cycle can exist in graph\\
3. Test if Hamiltonian cycle was found\\
If Hamiltonian cycle was not found:\\
4. Visit edge $a_{x_3} \rightarrow  a_{x_4}$\\
5. Remove unnecessary edges \dots \\
\vdots \\
\\
${x_{1}, x_{2}, \dots , x_{N} \in \{{1, 2, \dots , M}}\}$
\\\\
\\Algorithm doesn't visit a path by visiting one vertex after another.\\
\\Algorithm doesn't necessarily need to visit every edge in path to know if it is not a Hamiltonian cycle or if it is a Hamiltonian cycle.
\begin{figure}[h!]
    \centering
    \includegraphics[width=0.49\textwidth]{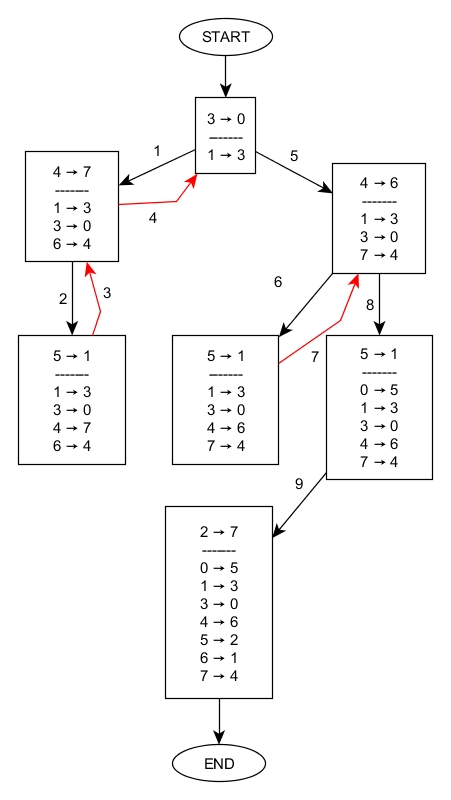}
    \caption{Example of usage of algorithm}
\end{figure}
\newpage
\subsection{Correct order}
Let's consider graph $G$. $G$ contains vertices: $2$ and $3$, vertex $2$ has $5$ neighbours and vertex $3$ has $3$ neighbours. Graph $G$ has only one Hamiltonian cycle. \\
Algorithm will test first vertex $3$ because it will give algorithm probability equal to $1/3$ of choosing edge that is in Hamiltonian cycle, which is better than probability equal to $1/5$. \\ \\
\begin{figure}[h!]
    \centering
    \includegraphics[width=0.75\textwidth]{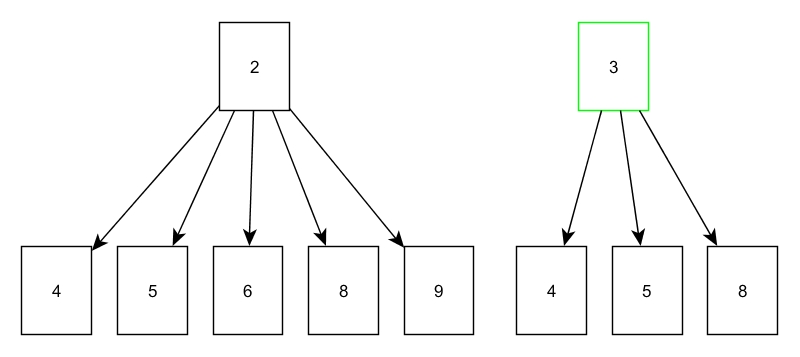}
    \caption{Algorithm will test first vertex $3$}
\end{figure}
\noindent
\\
\textbf{When algorithm decide which vertex should it test first, it will decide to choose vertex with smaller degree.}
\\ \\
Algorithm creates \textbf{correct order}, it is a list of vertices in graph ordered by their degrees in ascending order.
\subsection{Start vertex}
\label{53}
Algorithm starts the search in graph with:
\begin{enumerate}
\item first vertex in correct order from graph with degree greater than 1, if such a vertex doesn't exist it will start with first vertex in graph
\item first vertices in correct order
\end{enumerate}
\subsection{Next edge}
\label{54}
When algorithm tests vertex $A$, it have to decide which of $A$'s edges should it test first. \\ Let's consider following situation: vertex $A$ has 3 neighbours: $N, M$ and $P$, their degrees in opposite graph are: $N - 5, M - 4$ and $P-8$. Algorithm will test first edge $A \rightarrow M$ because it will give algorithm probability equal to $1/4$ of choosing edge that is in Hamiltonian cycle. To check edges in such an order, adjacency list of currently visited vertex is sorted by correct order from opposite graph. 
\subsection{Next vertex}
\label{55}
Next visited vertex is selected:
\begin{enumerate}
\item   as in \ref{53}.1
\item second vertex in currently tested edge - as in brute-force search
\end{enumerate}

\section{Features}
\begin{enumerate}
\item Attempt of removal of unnecessary edges and testing if Hamiltonian cycle can exist in graph occurs:
\begin{enumerate}
\item \label{61a} before search for Hamiltonian cycle begins.
\item \label{61b} with every recursive call of function that searches for Hamiltonian cycle - with visiting every edge in tested path.
\end{enumerate}
\item \label{62} When algorithm makes the decision to test edge $A \rightarrow X$ it removes all of the other edges from $A$.
\item \label{63} Edges removed with choosing the wrong edge are restored.
\item \label{64} Edges removed with choosing the wrong path are restored.
\end{enumerate}
\section{Algorithm}
\begin{algorithm}
\caption{Algorithm}
\begin{algorithmic}
\STATE START
\STATE Remove multiple edges and loops
\STATE Dirac's theorem test
\FOR{$stage=1$ to $8$}
\STATE Initialize variables
\STATE Analyze graph
\IF{Hamiltonian cycle can't exist}
\STATE END, Answer = "NO"
\ENDIF
\STATE FindHamiltonianCycle(\ref{53})
\IF{Answer was found}
\STATE END
\ENDIF
\ENDFOR
\end{algorithmic}
\end{algorithm}
\subsection{Stages}
\label{Stages}
$N$ - number of vertices in tested graph. Except for stage 6, algorithm uses \ref{55}.1.
\\
\\ Stage 0:
\begin{enumerate}
\item Check the closure of graph using Dirac's theorem to see if the graph is a Hamiltonian - \ref{DiracsTheorem}. 
\end{enumerate}
 Stage 1:
\begin{enumerate}
\item Graph analysis using \textbf{basic rules}(\ref{UN} \ref{NotEnoughUN} \ref{SingleEdge} \ref{SinglePath} \ref{Vertices1} \ref{VerticesMoreThan1} \ref{Alley} \ref{GraphDisconnected}  \ref{Never1HC_TLD} \ref{Never1HC_SE} \ref{Never1HC_RE}).
\item Check $2 * N$ paths by DFS in graph, using basic rules.
\end{enumerate}
Stage 2:
\begin{enumerate}
\item For "fully undirected" graphs only - analysis of closure of graph with basic rules. 
\end{enumerate}
Stage 3:
\begin{enumerate}
\item Graph analysis using basic rules and:
\begin{enumerate}
\item \ref{addMethod}
\item \ref{LN}
\item \ref{ALN}
\item \ref{Never1HC_FECR} - used only for "not very irregular" and "very undirected" graphs.
\end{enumerate}
\item Check $4 * N$ paths by DFS in graph, using basic rules.
\end{enumerate}
Stage 4:
\begin{enumerate}
\item \ref{CE}
\item Graph analysis using basic rules and:
\begin{enumerate}
\item \ref{LN}
\item \ref{ALN}
\item \ref{FECR} - used only for "not very irregular" and "very undirected" graphs.
\end{enumerate}
\item Check up to ${N}^{2}$ paths by DFS in graph, using basic rules.
\end{enumerate}
Stage 5:
\begin{enumerate}
\item Graph analysis using basic rules and:
\begin{enumerate}
\item \ref{addMethod}
\item \ref{LN}
\item \ref{ALN}
\item \ref{LNE}
\end{enumerate}
\item Check $2 * N$ paths by DFS in closure of graph, using basic rules.
\end{enumerate}
Stage 6:
\begin{enumerate}
\item Check $2 * N$ paths by DFS in graph, using basic rules, using \ref{55}.2.
\end{enumerate}
Stage 7:
\begin{enumerate}
\item Check $N$ paths by DFS in the graph, starting from $N/2$ different vertices - \ref{53}.2, using basic rules. 
\end{enumerate}
Stage 8:
\begin{enumerate}
\item Graph analysis using basic rules and:
\begin{enumerate}
\item \ref{addMethod}
\item \ref{combination}
\item \ref{LN}
\item \ref{ALN}
\item \ref{LNE}
\item \ref{recursion}
\end{enumerate}
\item Check $2 * N$ paths by DFS in closure of graph, using basic rules and:
\begin{enumerate}
\item \ref{addMethod}
\item \ref{LNE}
\end{enumerate}
\end{enumerate}

The algorithm deliberately does not use all its capabilities at once, because for the vast majority of tested graphs, it would only unnecessarily extend the time of determining whether the graph has a Hamiltonian cycle. 
\\
\\ If the algorithm is unable to determine whether the graph has a Hamiltonian cycle using the above-mentioned stages, it returns the information that it has not found an answer. 
\newline The above-described limits of the paths to be checked exist so that the algorithm has polynomial complexity. 
\begin{algorithm}
\caption{FindHamiltonianCycle(Vertex A)}
\begin{algorithmic}
\STATE $level \leftarrow level+1$
\IF{Hamiltonian cycle was found}
\STATE Answer = "Hamiltonian cycle was found"
\ENDIF
\STATE Sort neighbours of A //\ref{54}
\FOR{$i=0;i<A's \ degree;i \leftarrow i+1$}
\STATE B $\leftarrow$ A's i-th neighbour
\STATE C $\leftarrow$ current graph
\STATE Remove edge A $\rightarrow$ B
\STATE D $\leftarrow$ edges: A $\rightarrow$ all neighbours of A in current graph
\STATE Remove edges D //8.\ref{62}
\STATE Add edge A $\rightarrow$ B
\STATE AnalyzeGraph(D)  //8.\ref{61b}
\IF{Hamiltonian cycle can exist}
\STATE FindHamiltonianCycle(\ref{55})
\STATE Restore graph to C    //8.\ref{64}
\ELSE
\STATE Add edges D   //8.\ref{63}
\ENDIF
\ENDFOR
\STATE $level \leftarrow level-1$
\IF{$level==0$}
\STATE Answer = "Hamiltonian cycle doesn't exist in graph"
\ENDIF
\end{algorithmic}
\end{algorithm}
\subsection{Which vertices should be analyzed?}
Before search for Hamiltonian cycle begins every vertex is analyzed.
\\ 
In every recursive call of function that searches for Hamiltonian cycle two types of vertices should be analyzed:
\begin{enumerate}
\item \label{921} vertices whose adjacency list where changed in current call of function
\item vertices whose adjacency list are supersets of any of adjacency list of \ref{921}
\end{enumerate}
\section{Examination of algorithm}
Algorithm was tested on 4 types of graphs:
\begin{itemize}
\item "directed regular"
\item "directed irregular"
\item "undirected regular"
\item "undirected irregular''
\end{itemize}
Graph "undirected" is a graph which has many edges $X \rightarrow Y$ and also $Y \rightarrow X$.\\
Graph "irregular" is a graph with median of all vertices degrees being much different than degree of vertex with highest number of neighbours.
\\
\\
\begin{figure}[h!]
    \centering
    \includegraphics[width=0.79\textwidth]{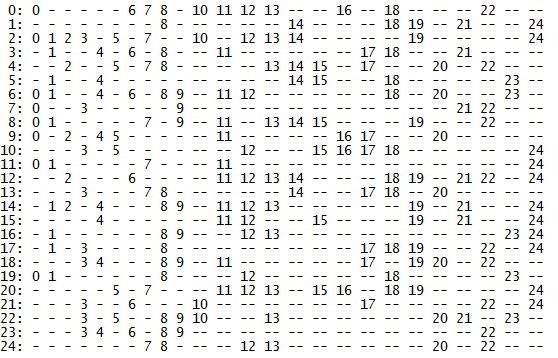}
    \caption{Example of "directed regular" graph}
\end{figure}
\begin{figure}[h!]
    \centering
    \includegraphics[width=0.79\textwidth]{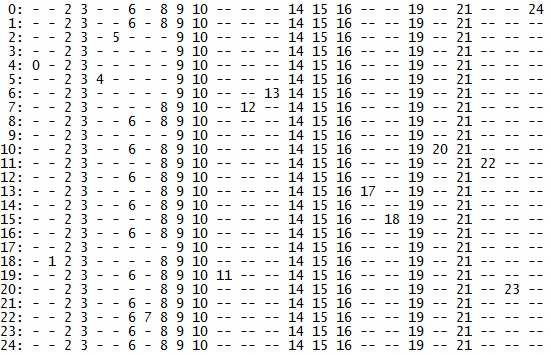}
    \caption{Example of "directed irregular" graph}
\end{figure}
\begin{figure}[h!]
    \centering
    \includegraphics[width=0.95\textwidth]{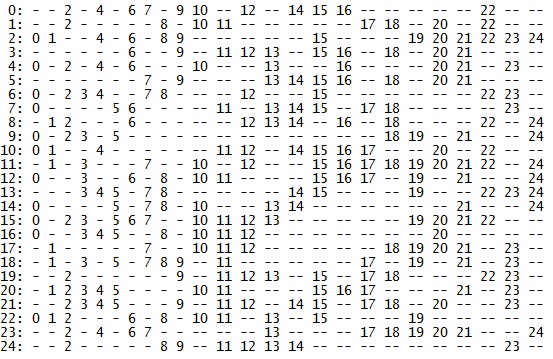}
    \caption{Example of "undirected regular" graph}
\end{figure}
\begin{figure}[h!]
    \centering
    \includegraphics[width=0.95\textwidth]{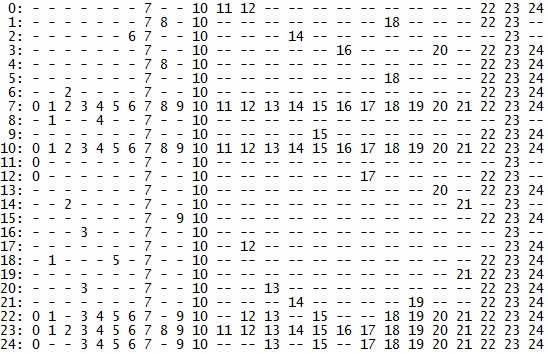}
    \caption{Example of "undirected irregular" graph}
\end{figure}
\newpage
\newpage
\hspace{10cm}
\subsection{Algorithms correctness}
Algorithms correctness was checked by passing the algorithm graphs with Hamiltonian cycle and testing if algorithm would confirm existence of Hamiltonian cycle. Graphs with 50 vertices were tested.
\\ Tested graphs can be downloaded from:\\
\url{http://figshare.com/articles/Correctness_Test/1057640}.
\\ 10 000 "directed regular" graphs are located in directory "CT\_50\_T\_T".
\\ 10 000 "directed irregular" graphs are located in directory "CT\_50\_T\_F".
\\ 10 000 "undirected regular" graphs are located in directory "CT\_50\_F\_T".
\\ 10 000 "undirected irregular" graphs are located in directory "CT\_50\_F\_F".
\\For every tested graph, algorithm confirmed existence of Hamiltonian cycle.

\subsection{Algorithms efficiency}
Following results were given on AMD FX-8350 4.00 Ghz CPU, only on a single CPU core, on Windows 7. Algorithm was tested with every one of 4 types of graphs. 
\\I tested the algorithm performance on a total of 10 000 000 graphs: 
\begin{enumerate}
\item 4 000 000 graphs with 25 vertices, in a graph with 25 vertices there can be a maximum of 600 edges.
\item 3 000 000 graphs with 50 vertices, in a graph with 50 vertices there can be a maximum of 2 450 edges.
\item 2 000 000 graphs with 75 vertices,  in a graph with 75 vertices there can be a maximum of 5 550 edges
\item 1 000 000 graphs with 100 vertices, in a graph with 100 vertices there can be a maximum of 9 900 edges.
\end{enumerate}

\textbf{The use of only a part of the algorithm's capabilities - described as stage 0 and 1, made it possible to determine whether the graph has a Hamiltonian cycle for 9 997 927 graphs, which is over 99.97\% of all examined graphs.}

The time to examine graphs in stages 0 and 1 took up to:
\begin{enumerate}
\item for graphs with 25 vertices - 0.05 seconds
\item for graphs with 50 vertices - 0.20 seconds
\item for graphs with 75 vertices - 0.95 seconds
\item for graphs with 100 vertices - 2.07 seconds 
\end{enumerate}
\noindent
The number of graphs that required the use of full capabilities of the algorithm was 2 073. Among these graphs, the algorithm was able to determine whether the graph has a Hamiltonian cycle for 2 051 graphs in time:
\begin{enumerate}
\item for graphs with 25 vertices - in an average time of 0.063 seconds and maximum time of 0.33 seconds
\item for graphs with 50 vertices - in an average time of 0.33 seconds and a maximum time of 5.60 seconds
\item for graphs with 75 vertices - in an average time of 1.17 seconds and maximum time of 19.52 seconds
\item for graphs with 100 vertices - in an average time of 2.52 seconds and maximum time of 33.4 seconds 
\end{enumerate}
Only for 22 graphs, the algorithm was unable to find a solution in described above stages - meaning, in a relatively short time. 
\newline
\newline
\noindent
 ''Directed regular'' graphs are located in directory "E\_A\_T\_T".\\
 "Directed irregular" graphs are located in directory "E\_A\_T\_F".\\
 "Undirected regular" graphs are located in directory "E\_A\_F\_T".\\
 "Undirected irregular" graphs are located in directory "E\_A\_F\_F"\\
where A is number of vertices.\\ \\
All tested graphs can be downloaded from:\\
\url{http://figshare.com/authors/Pawe_Kaftan/568545}.
\newline2 073 graphs that needed more than basic capabilities of algorithm, can be downloaded from:\\
\url{https://figshare.com/articles/dataset/Graphs_not_solved_in_stages_0_1_v6_zip/20562648}.
\newline22 graphs that algorithm was unable to find a solution, can be downloaded from:\\
\url{https://figshare.com/articles/dataset/Graphs_not_solved_in_all_stages_v6_zip/20562600}.
\subsection{How the examined graphs look like}
\begin{figure}[h!]
    \centering
    \includegraphics[width=0.8\textwidth]{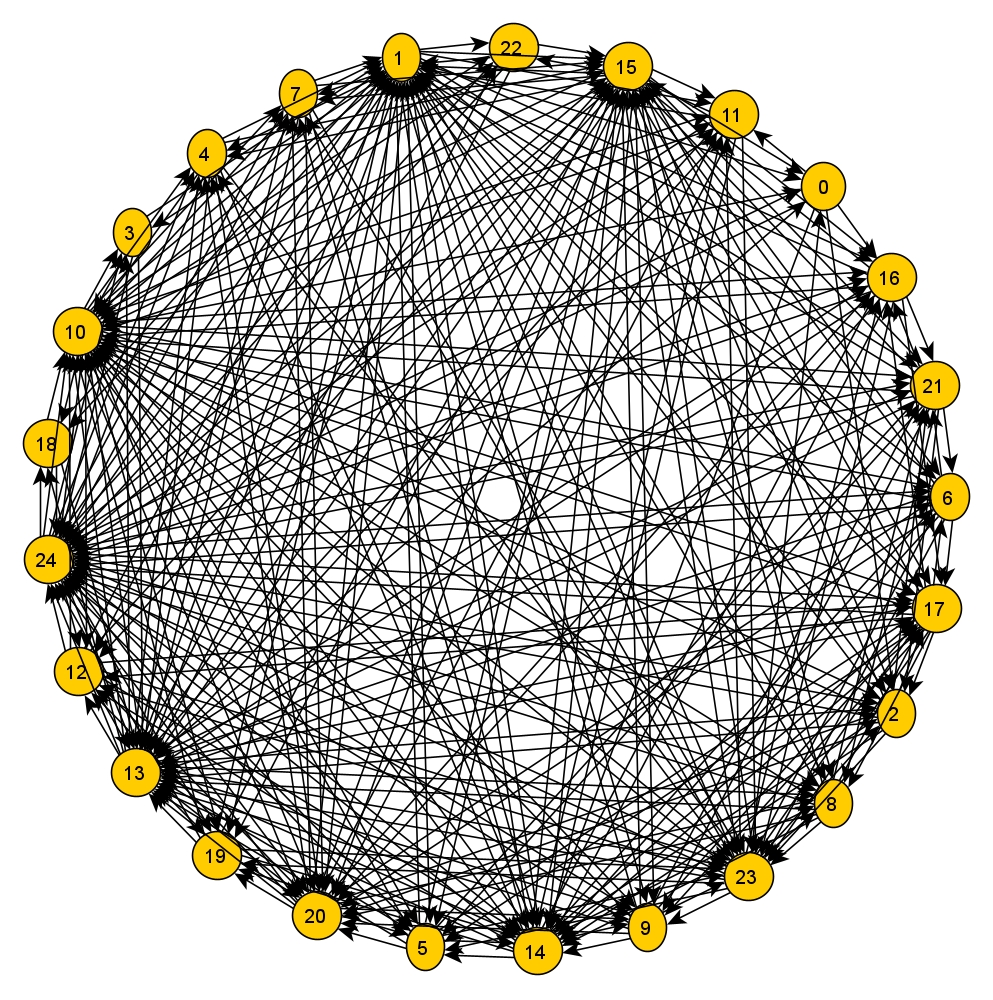}
    \caption{Example of graph with 25 vertices and 313 edges}
\end{figure}
\begin{figure}[h!]
    \centering
    \includegraphics[width=0.625\textwidth]{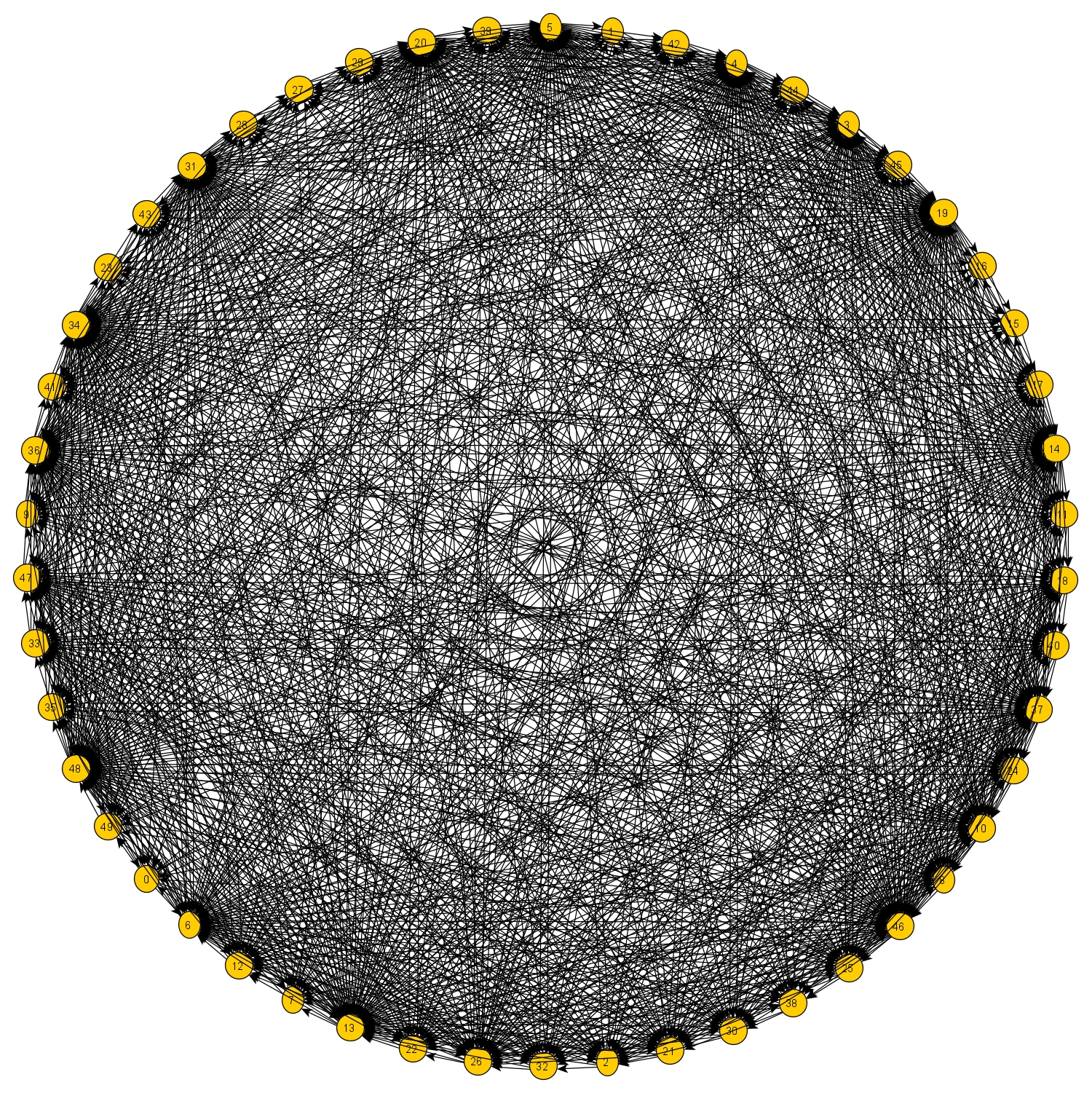}
    \caption{Example of graph with 50 vertices and 1246 edges}
\end{figure}
\begin{figure}[h!]
    \centering
    \includegraphics[width=0.625\textwidth]{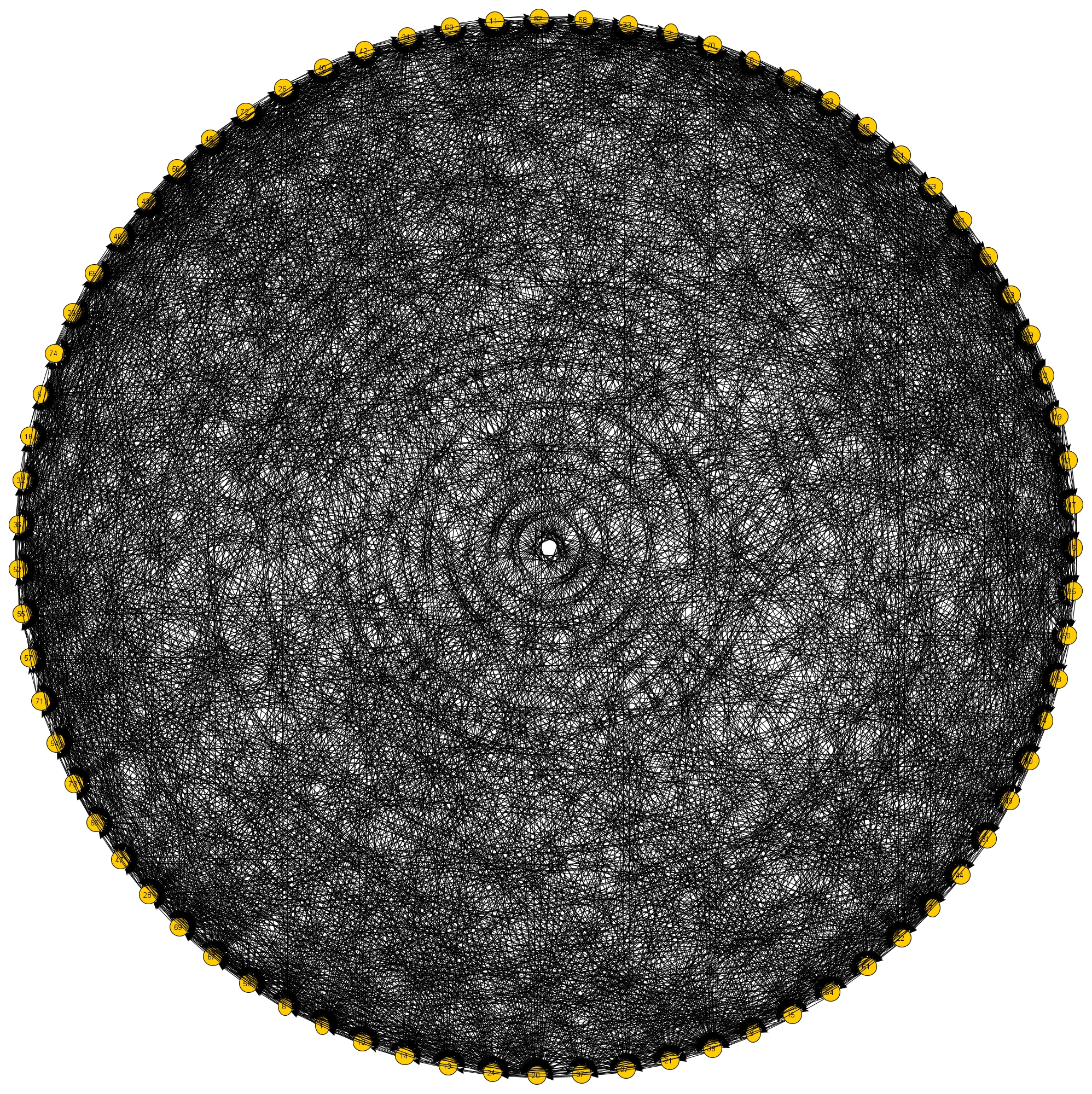}
    \caption{Example of graph with 75 vertices and 2824 edges}
\end{figure}
\begin{figure}[h!]
    \centering
    \includegraphics[width=0.95\textwidth]{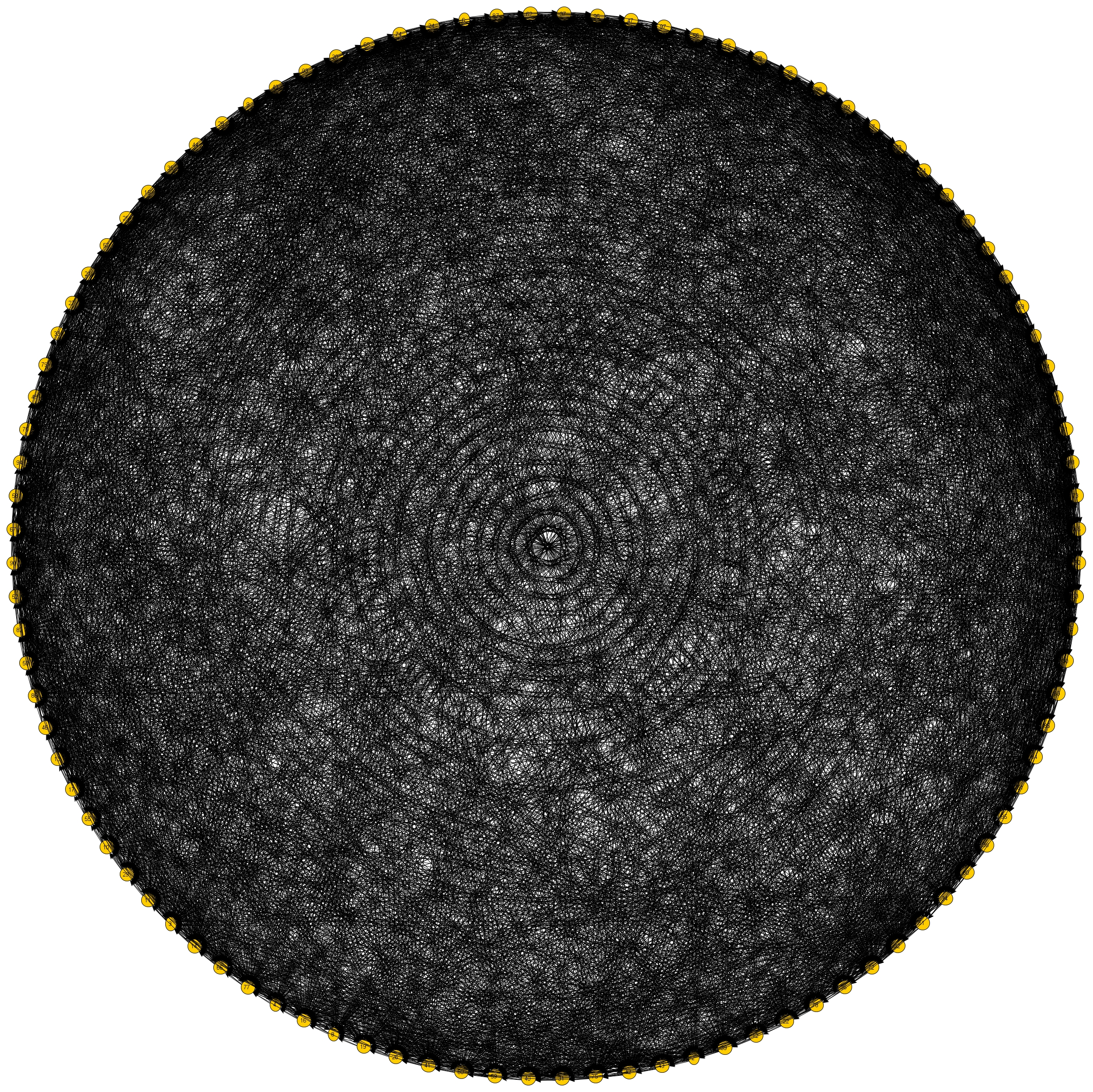}
    \caption{Example of graph with 100 vertices and 5020 edges}
\end{figure}
\newpage

\section{Algorithm implementation}
Algorithm implementation in C\#:\\ \url{https://figshare.com/articles/software/Algorithm_v6_zip/20563158}
\section{Contact}
\url{https://pl.linkedin.com/in/pawe%C5%82-kaftan-99b27919a}
\end{document}